\documentclass[journal]{IEEEtran}
\usepackage{amsthm}

\newtheorem{lemma}{Lemma}
\newtheorem{theo}{Theorem}
\newtheorem*{proposition}{Proposition}

\newtheorem{cor}{Corollary}
\usepackage{amsfonts, amssymb}

\usepackage{cite}
\usepackage{color}
\usepackage{graphicx}
\usepackage[cmex10]{amsmath}

\usepackage{array}
\usepackage{mdwmath}

\usepackage[tight,footnotesize]{subfigure}

\begin{document}
%
\title{On Conditions for Linearity of Optimal Estimation }

\author{Emrah~Akyol,~\IEEEmembership{Student Member,~IEEE,}
        Kumar~Viswanatha,~\IEEEmembership{Student Member,~IEEE,}
        and~Kenneth~Rose,~\IEEEmembership{Fellow,~IEEE}
\thanks{Authors are with the Department
of Electrical and Computer Engineering, University of California, Santa Barbara,
CA, 93106 USA e-mail: \{eakyol, kumar, rose\} @ece.ucsb.edu}

\thanks{The material in this paper was presented
in part at the IEEE Information Theory Workshop (ITW), Dublin, Aug 2010 and IEEE Statistical Signal Processing Workshop (SSP), Nice, France, June 2011.}

\thanks{Copyright (c) 2011 IEEE. Personal use of this material is permitted.  However, permission to use this material for any other purposes must be obtained from the IEEE by sending a request to pubs-permissions@ieee.org.}
}

\markboth{IEEE Transactions  on Information Theory,~Vol.~xx, No.~xx, xx~xx}%
{Akyol \MakeLowercase{\textit{et al.}}: On Conditions for Linearity of Optimal Estimation}

\maketitle

\begin{abstract}
When is optimal estimation linear? It is well known that, when a Gaussian source is contaminated with Gaussian noise, a linear estimator minimizes the mean square estimation error. This paper analyzes, more generally, the conditions for linearity of optimal estimators. Given a noise (or source) distribution, and a specified signal to noise ratio (SNR), we derive conditions for existence and uniqueness of a source (or noise) distribution for which the $L_p$ optimal estimator is linear. We then show that, if the noise and source variances are equal, then the matching source must be distributed identically to the noise. Moreover, we prove that the Gaussian source-channel pair is unique in the sense that it is the only source-channel pair for which the mean square error (MSE) optimal estimator is linear at more than one SNR values. Further, we show the asymptotic linearity of MSE optimal estimators for low SNR if the channel is Gaussian regardless of the source and, vice versa, for high SNR if the source is Gaussian regardless of the channel. The extension to the vector case is also considered where besides the conditions inherited from the scalar case, additional constraints must be satisfied to ensure linearity of the optimal estimator.
\end{abstract}


%
\IEEEpeerreviewmaketitle

\section{Introduction}
%
%
%
%
\IEEEPARstart{C}{onsider} a basic problem in estimation theory, namely, source estimation from a signal received through a channel with additive noise, given the statistics of both source and channel. The optimal estimator that minimizes the mean square error (MSE) is usually a nonlinear  function of the observation. A frequently exploited result in estimation theory concerns the special case of Gaussian source and Gaussian noise, a case in which the MSE optimal estimator is guaranteed to be linear. An open follow-up question considers the existence of other cases exhibiting such a ``coincidence'', and more generally the characterization of conditions for linearity of optimal estimators for general distortion measures.

This problem also has practical importance beyond theoretical interest, mainly due to significant complexity issues in both design and operation of estimators. Specifically, the optimal estimator generally involves entire probability distributions, whereas linear estimators require only up to second-order statistics for their design.  Moreover, unlike the optimal estimator which can be an arbitrarily complex function that is difficult to implement, the linear estimator consists of a simple matrix-vector operation. Hence, linear estimators are more prevalent in practice, despite their suboptimal performance in general. They also represent a significant temptation to ``assume'' that processes are Gaussian, sometimes despite overwhelming evidence to the contrary. Results in this paper identify the cases where a linear estimator is optimal, and when the use of linear estimators is justified in practice without recourse to complexity arguments.

The estimation problem in general has been studied intensively in the literature \cite{allen1938theorem,rothschild1947lois, rao2005probability, olkin, billingsley2008probability,rao1947note}. Our preliminary results appeared in \cite{estimation_itw,akyol2011multidimensional}. It is known that, for stable distributions\footnote{A distribution is called stable if for independent identically distributed $ X_1, X_2, X $;  for any constants $a$, $b$; the random variable $aX_1 + bX_2$ has the same distribution as $cX + d$ for some constants $c$ and $d$ \cite{billingsley2008probability}.} (which includes the Gaussian distribution as the only finite variance member), the optimal estimator is linear at all signal to noise ratios (SNR). Stable distributions are a subset of a family called infinitely divisible distributions which, as we show in this paper, satisfy the derived necessary conditions for the existence of a matching source/noise distribution such that the optimal estimator is linear at any SNR level. Our main contribution relative to prior work, which studied linearity as it applies simultaneously at all SNR levels, focuses on the linearity of optimal estimation for the $L_p$ norm and its dependence on the SNR level. Specifically, we present the optimality conditions for linearity of optimal estimators at a specified SNR, where optimality is in the sense of the $L_p$ norm. As an important special case, we investigate the $p=2$ case (mean square error) in detail. Note that a similar problem has been studied in \cite{laha,balakrishnan} for the special case of the mean square error, albeit without further study related to questions of existence and uniqueness of ``matching" distributions. We show that the necessary conditions presented in \cite{laha,balakrishnan} are subsumed in our general necessary and sufficient conditions; and specify conditions for which such matching distributions exist and are unique. The analysis is then extended to the case of vector spaces. Interestingly, this extension is non-trivial and new constraints, beyond those inherited from the scalar case, must be satisfied to ensure linearity of optimal estimation.

Five results are provided on the linearity of optimal estimation.  First, we show that if a given noise (alternatively, a given source) distribution satisfies certain conditions, there always exists a matching source (alternatively, noise) distribution of a given power, for which the optimal estimator is linear. We further identify conditions under which such a matching distribution does {\it{not}} exist. Secondly, we show that if the source and the noise have the same variance, they {\it{must}} be identically distributed to ensure the linearity of the optimal estimator.
Having established more general conditions for linearity of optimal estimation, one wonders in what precise sense the Gaussian case may be special. This question is answered by the third result. We consider the optimality of linear estimation at multiple SNR values. Let random variables $X$ and $Z$ be source and noise, respectively, and allow for scaling of either to produce varying levels of SNR. We show that if the optimal estimator is linear at more than one SNR value, then both the source $X$ and the noise $Z$ must be Gaussian. In other words, the Gaussian source-noise pair is unique in the sense that it offers linearity of optimal estimators at multiple SNR values (in fact the optimal estimator is linear at all SNR as is well known).
As a fourth result, we show that the MSE optimal estimator converges to a linear estimator for any source and Gaussian noise at asymptotically low SNR, and vice versa, for any noise and Gaussian source at asymptotically high SNR.

Finally, we analyze the vector case, where conditions for linearity of optimal estimation are more stringent. We show that for a vector source-channel pair with identical dimensions, the conditions derived for the scalar case become necessary conditions in a transform domain, where the transform jointly diagonalizes the source and channel covariance matrices. We further derive the additional, complementary conditions that must be satisfied to achieve sufficiency.

The paper is organized as follows: we review optimal and linear estimation in Section II, present the main result in Section III, its main corollaries in Section IV, the vector case in Section V,  and conclusions in Section VI.

 \section{Review of Optimal and Linear Estimation}
\subsection {Preliminaries and Notation}
\begin{figure}
\centering
\includegraphics[scale=0.2]{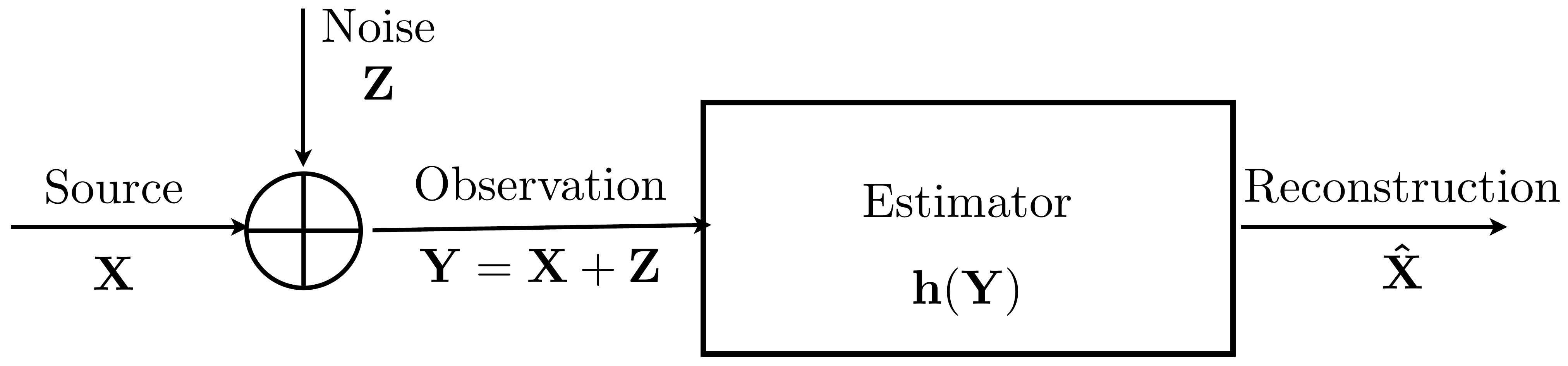}
\label{figure}
 \caption{The general setup of the problem}
\end{figure}
Let $\mathbb R$, $\mathbb R^+$, and $\mathbb N$  denote the respective sets of real numbers, positive real numbers and natural numbers. In
general, lowercase letters (e.g., $x$) denote scalars, boldface lowercase (e.g., $\boldsymbol x$) vectors, uppercase (e.g., $U, X$) matrices and random variables, and boldface uppercase (e.g., $\boldsymbol X$) random
vectors. Unless otherwise specified, vectors and random vectors have length $m$, and matrices
have size $m \times m$. The $k^{th}$ element of vector $\boldsymbol x$ is denoted by $[\boldsymbol x]_k$ and the ($i$, $j$)-th
element and the $k^{th}$ column of the matrix $U$ by $[U]_{ij}$ and ${[U]_{k}}$ respectively. $U^{-T}$ denotes $(U^{T})^{-1}$. $\mathbb E[\cdot]$,  $ R_X$, and $  R_{XZ}$ denote the expectation, covariance of $\boldsymbol X $ and cross covariance of $\boldsymbol X$ and $\boldsymbol Z$ respectively. $\nabla$ denotes the gradient and $\nabla_x$ denotes the partial gradient with respect to $\boldsymbol x$. $F^{(k)} (\cdot)$ denotes the $k^{th}$ order derivative of the function $F(\cdot)$, i.e., $F^{(k)} (x)= \frac{d^k F(x)}{d x^k}$.

We consider the problem of estimating source $X$ given the observation $Y=X+Z$, where $X$ and $Z$ are independent, as shown in Figure 1. Let $X$ and $Z$ be scalar zero mean\footnote{The zero mean assumption is not crucial, but it considerably simpliÞes the notation. Therefore, it is kept throughout the paper.} random variables with respective densities $f_X(\cdot)$ and $f_Z(\cdot)$ and  characteristic  functions $F_X(\omega)$ and $F_Z(\omega)$. A density $f(x)$ is said to be symmetric if it has an even characteristic function\footnote{Note that this definition requires generalization to symmetry about the mean when one drops the assumption of  zero-mean random variables.}: $f(x)=f(-x)$ $ \forall x \in \mathbb R $. The SNR is $\gamma=\frac {\sigma_x^2}{\sigma_z^2}$, where $\sigma_x^2=\mathbb E\{X^2\}$ and  $\sigma_z^2=\mathbb E\{Z^2\}$. In any statement concerning $L_p$ norm, all random variables are assumed to have finite $p^{th}$ order moments, e.g., in any result associated with MSE we assume finite variances, $\sigma_x^2<\infty, \sigma_z^2<\infty$. All the logarithms in the paper are natural logarithms and may in general be complex.

In the rest of this section, we review and derive some preliminary results concerning optimal estimators which will be useful in the following sections in proving our main results. An estimator $h(\cdot)$ is a  function of the observation and is said to be optimal if it minimizes the cost  functional
\begin{equation}
J(h)=\mathbb E \left \{ \Phi(X,h(Y))\right \}
\label{mse}
\end{equation}
for a given distortion measure $\Phi$, which is assumed to be first order differentiable. Specializing (\ref{mse}) to a difference distortion measure, we explicitly  get:
\begin{equation}
J(h)=\int \int  \Phi (x-h(y)) f_X(x) f_{Z}(y-x) dx dy
\label{mse2}
\end{equation}

 To obtain the necessary conditions for optimality,  we apply the standard method in variational calculus \cite{Luenberger}:
\begin{equation}
\frac{\partial}{\partial \epsilon}  J\left (h+\epsilon \eta\right ) \bigg|_{\epsilon =0} =0
\label{general}
\end {equation}
for all variation  functions $\eta(\cdot)$. Then, (\ref{general}) yields
\begin{equation}
\int \int  \Phi' (x-h(y)) \eta(y) f_X(x) f_{Z}(y-x) dx dy =0
\label{mse3}
\end{equation}
or,
\begin{equation}
\mathbb E \left \{ [\Phi' (X-h(Y)] \eta(Y)\right \}=0
\label{mse4}
\end{equation}
for all  variation  functions $\eta(\cdot)$, where $\Phi '$ is the derivative of $\Phi$.
\subsection {Optimality condition for $L_p$ norm}
Hereafter, we will specialize to the case of the $L_p$ metric with $p=2 \rho, \text{ } \rho \in \mathbb N$, i.e., $\Phi(x)=|x|^p$ for even\footnote{Although some of the high level results may be derived for all natural p, in this paper we focus on even p which enables considerable simplification of the results, hence providing much insight and clear intuitive interpretation of the solution.}  and natural $p$. Using the fact that $\frac{d}{dx} |x|^p =p\frac{|x|^p}{x},  \forall x \in {\mathbb R} -\{0\}$, we derive the necessary condition for optimality of an estimator as :

\begin{equation}
\mathbb E \left \{  [X-h(Y)]^{p-1} \eta(Y)\right \}=0 
\label{mse5_2}
\end{equation}

  Note that for $p=2$, or $\Phi(x)=x^2$, this condition reduces to the well known orthogonality condition of MSE, i.e., the following holds :
  \begin{equation}
  \mathbb E \left \{ [(X-h(Y)] \eta(Y)\right \}=0
  \label{orthogonality}
   \end{equation}
   for any $\eta(\cdot)$  function. The MSE optimal estimator $h(Y)=\mathbb E\left \{ X|Y\right \}$ can be directly obtained from (\ref{orthogonality}). The following lemma formally states that the above necessary condition, (\ref{mse5_2}), is also sufficient for minimizing $L_p$ norm.

\begin{lemma}
\label{lemma1}
The necessary condition stated in (\ref{mse5_2}) is sufficient. Moreover, the optimal estimator is unique almost everywhere (optimal estimators may only differ over a set of zero measure).
\end{lemma}
\begin{IEEEproof} See Appendix \ref{appa}. \end{IEEEproof}

\subsection {$L_p$ Optimal Linear Estimation}
To derive the optimal linear estimator, the variation  function $\eta(y)$ must be made linear to ensure linearity of $h(y)+\epsilon \eta(y)$. Plugging $h(Y)=kY$ and $\eta(Y)=aY$ (for some $a \in \mathbb R$) in (\ref{mse5_2}) and omitting straightforward steps, we obtain the condition for optimal linear estimation to be:
\begin{equation}
\mathbb E  \left \{ (X-kY)^{p-1} Y \right \} =0
\label{optlinear}
\end{equation}
\noindent The optimal scaling coefficient $k$ can be found by plugging $Y=X+Z$ into (\ref{optlinear}). Observe that for $p=2$, we get the well known result $k=\frac{\gamma}{\gamma+1}$.

\subsection{Gaussian Source and Channel}
We next consider the special case in which both $X$ and $Z$ are Gaussian, $X\sim \mathcal N(0,\sigma_x^2)$ and $Z \sim \mathcal  N(0,\sigma_z^2)$. The linear estimator
\begin{equation}
\label{lin-est}
h(Y)= \frac{\gamma}{\gamma+1}Y
\end{equation}
 is well known to be the optimal MSE estimator. A relatively less known fact is that this linear estimator is optimal more generally for the $L_p$ norm \cite{sherman}. It is straightforward to show that this linear estimator satisfies (\ref{mse5_2}) by rendering the reconstruction error $X-h(Y)$ independent of $Y$.

\section{Conditions for Linearity of Optimal Estimation}

In this section,  we find the necessary and sufficient conditions in terms of characteristic functions $F_X(\omega)$ and $F_Z(\omega)$ that ensure that $h(Y)=k Y$ is the optimal estimator for some $k \in \mathbb R$. We first provide the result for the $L_p$ norm, which takes the form of a differential equation that must be satisfied to ensure linearity of optimal estimation, and then specialize it to the MSE case. 
\subsection{$L_p$ Norm Condition}
As stated previously for any  $L_p$ norm result, the characteristic functions of the source and noise $F_X(\omega)$ and $F_Z(\omega)$ are assumed to be $p^{th}$ order differentiable. 
 \begin{theo}
 \label{firsttheo}
Given an $L_p$ distortion measure, source $X$ and noise $Z$ with characteristic functions $F_X(\omega)$ and $F_Z(\omega)$ respectively,  the optimal estimator is linear, $h(Y)=kY$, where $Y=X+Z$, {\em if and only if} the following differential equation is satisfied:
\begin{align}
\label{main_Lp}
\sum_{m=0}^{p-1} \left({p-1\atop m}\right) F_X^{(m)}(\omega)F_Z^{(p-1-m)}(\omega)\left(\frac{k-1}{k}\right)^{m} = 0 \, \,
\end{align}
\end{theo}
\begin{IEEEproof}
See Appendix \ref{appb}.
\end{IEEEproof}

\subsection{Specializing to MSE: The Matching Condition}

In this section, we explore the impact of Theorem \ref{firsttheo} for the special case of the mean square error distortion metric, i.e., $p=2$. More precisely, we wish to find the entire set of source and channel distributions such that $h(Y)=\frac{\gamma}{\gamma+1} Y$ is the optimal estimator for a given SNR, $\gamma$. Note that this condition was derived, in another context \cite{laha,balakrishnan}, albeit without consideration of important implications which we focus on, including the conditions for existence and uniqueness of matching distributions. Specifically, we identify the conditions for existence and uniqueness of a source distribution that \textit{matches} the noise (and vice versa) in a way that guarantees the linearity of the optimal estimator. We state the main result for MSE in the following theorem.
\begin{theo}
Given SNR level $\gamma$, and noise $Z$ with  characteristic function $F_Z(\omega)$, there exists a source $X$ for which the optimal estimator is linear {\em if and only if} the function
$$F(\omega)= F_Z^{\gamma}(\omega)$$
is a legitimate characteristic function. Moreover, if $F(\omega)$ is legitimate, then it is the characteristic function of the matching source, i.e., $F_X(\omega)=F(\omega)$.

(An equivalent theorem holds where we replace ``noise'' for ``source'' everywhere, i.e., given source and SNR level, we have a condition for existence of a matching noise.)
\end{theo}

\begin{IEEEproof}
Plugging $p=2$ and $k=\frac{\gamma}{\gamma+1}$ in (\ref{main_Lp}) yields
\begin{equation}
\label{above_eq3}
\frac{1}{F_X(\omega)}  \frac {dF_X(\omega)}{d\omega} =\gamma \frac{1}{F_Z(\omega)}  \frac {dF_Z(\omega)}{d \omega}
\end{equation}
or more compactly,
\begin{equation}
\label{above_eq5}
 \frac{ d}{d\omega} \log {F_X(\omega)}= \gamma \frac { d}{d\omega} \log {F_Z(\omega)}
\end{equation}
The solution to this differential equation is given by:
\begin{equation}
\label{above_eq6}
\log{F_X(\omega)}= \gamma \log{F_Z(\omega)}+C
\end{equation}
where $C$ is a constant. Imposing $F_Z(0)=F_X(0)=1$, we obtain $C=0$, which implies:
\begin{equation}
\label{main}
{F_X(\omega)}= F_Z^{\gamma}(\omega)
 \end{equation}
\end{IEEEproof}
Hence, given a noise distribution, the necessary and sufficient condition for the existence of a matching source distribution boils down to the requirement that $F_Z^{\gamma}(\omega)$ be a valid characteristic function. Moreover, if such a matching source exists, we have a recipe for deriving its distribution.

\subsection{Existence of a Matching Source for a Given Noise}
In this section, we study the conditions under which a matching source exists for a given noise distribution. During the course, we also study some important properties relating the matching distributions when they exist.

We begin with Bochner's theorem \cite{rao2005probability}, which states that a continuous function $F: \mathbb  R \rightarrow \mathbb C$ with $F(0)=1$ is a valid characteristic function if and only if it is \textit{positive semi-definite}.\footnote{Let $f: \mathbb  R \rightarrow \mathbb C$ be a complex-valued function, and $t_1,...,t_s$ be a set of points in $\mathbb  R$. Then $f$ is said to be positive semi-definite (non-negative definite) if for any $t_i \in \mathbb R$ and $a_i \in \mathbb  C$, $i=1,...,s$ we have
\begin{equation*}
\sum_{i=1}^{s}\sum_{j=1}^{s} a_i {a_j}^*f(t_i-t_j) \geq 0
\end{equation*}
where ${a_j}^*$ is the complex conjugate of $a_j$. Equivalently, we require that the $s\times s$ matrix constructed with $f(t_i-t_j)$ be positive semi-definite. If  function $f$ is positive semi-definite, its Fourier transform, is non-negative everywhere $F(\omega) \geq 0, \forall \omega \in \mathbb R$.  Hence, in the case of our candidate characteristic function, this requirement ensures that the corresponding density is indeed non-negative everywhere.}
Hence, the existence of a matching source depends on the positive semi-definiteness of $F_Z^\gamma(\omega)$.

We note that characterizing the entire set of $F_Z(\omega)$ where $F_Z^{\gamma}(\omega)$ is positive semi-definite is a long-standing open problem. Instead we illustrate the result with various cases of interest where $F_Z^{\gamma}(\omega)$ is, or is not, positive semi-definite.  Let us start with a simple but useful case.
\begin{cor}
\label{firstcorr}
If  SNR $\gamma \in \mathbb N$, a matching source distribution exists, regardless of the noise distribution.
\end{cor}
\begin{IEEEproof}
From (\ref{main}), natural $\gamma$ implies:
\begin{equation}
X=\sum_{i=1}^{\gamma} Z_i
\end{equation}
\noindent where $Z_i$ are independent and distributed identically to $Z$. Hence, $F_Z^\gamma(\omega)$ is a valid characteristic function and a matching $X$ exists.
\end{IEEEproof}

\noindent Next, we recall the concept of infinite divisibility, which is closely related to the problem at hand.

{\bf \noindent Definition \cite{lukacs1960characteristics}}: A distribution with characteristic function $F(\omega)$ is called infinitely divisible, if for each integer $k \geq 1$, there exists a characteristic function $F_k(\omega)$ such that
\begin{equation}
\label{infdiv}
F(\omega)=F_k^k(\omega)
\end{equation}
Alternatively, $f_X(\cdot)$ is infinitely divisible if and only if the random variable $X$ can be written for any $k$ as $X=\sum_{i=1}^k X_i $  where $\{X_i, i=1,..., k\}$ are independent and identically distributed.

Infinitely divisible distributions have been studied extensively in probability theory \cite{lukacs1960characteristics, steutel2003infinite}. It is known that Poisson, exponential, and geometric distributions as well as  the set of stable distributions (which includes the Gaussian distribution) are infinitely divisible. On the other hand, it is easy to see that distributions of discrete random variables with finite alphabets are not infinitely divisible.
\begin{cor}
\label{seccorr}
 A matching source distribution exists for all $\gamma \in \mathbb R^{+}$ {\it if and only if} $f_Z(\cdot)$ is infinitely divisible.
\end{cor}
\begin{IEEEproof}  We first note that if $f_Z(\cdot)$ is infinitely divisible, $F_Z^{1/j}(\omega)$ is a valid characteristic function for all natural $j$, as follows directly from the definition of infinite divisibility. Then, by Corollary \ref{firstcorr}, it follows that $F_Z^{i/j}(\omega)$ is also a valid characteristic function, which implies that so is $F_Z^r(\omega)$ for all positive rational $r>0$ since a rational $r$ means that $r=i/j$ for some natural $i$ and $j$. Using the fact that every $\gamma \in \mathbb R^{+}$ is a limit of a sequence of rational numbers $r_n$, and by the continuity theorem \cite{billingsley2008probability}, we conclude that $F_X(\omega)=F_Z^\gamma(\omega)$ is a valid characteristic function, and hence a matching source exists.

Towards showing the converse, note that if $F_X(\omega)=F_Z^{\gamma}(\omega)$ is a valid characteristic function for all $\gamma$, then $f_Z(\cdot)$ has to be infinitely divisible, because we can always choose $\gamma=\frac{1}{k}$ for $k \in \mathbb N $ and set $F_k(\omega)=F_X(\omega)$ in (\ref{infdiv}).
\end{IEEEproof}

However, note that at a given SNR, there may exist a matching source, even though $f_Z(\cdot)$ is not infinitely divisible. For example, a finite alphabet discrete random variable $V$ is not infinitely divisible but still can be $k$-divisible, where $k<|V|-1$ and $|V|$ is the cardinality of $V$. Hence, when $\gamma=\frac{1}{k}$, there may exist a matching source, even when the noise distribution is not infinitely divisible. Many examples follow directly from Corollary \ref{firstcorr}.

We next cite a theorem, regarding analytic characteristic functions, which will be useful in the proofs that follow.

Theorem \cite{lukacs1960characteristics}: A characteristic function $F(\omega)$ is analytic {\em if and only if} $F$ has finite moments of all orders and there exists a finite $\beta$ such that $\mathbb E\{|X^k|\} \leq k! \beta^k, \forall k\in \mathbb N$. This requirement is equivalent to the existence of a moment generating  function. A characteristic function $F(\omega)$ is analytic {\em if and only if} the moments $\mathbb E\{|X^k|\}$ uniquely characterize the distribution, which in general is not the case, see eg. \cite{shohat1943problem}.

A useful property of the matching pair, relating the analyticities of their characteristic functions is captured by the following corollary.
\begin{cor}
\label{analytic}
If $F_Z(\omega)$ (or  $F_X(\omega)$) is analytic, then the matching $F_X(\omega)$ (or  $F_Z(\omega)$), if it exists, is analytic.
\end{cor}
\begin{IEEEproof}
Recall the orthogonality property of the MSE optimal estimator (\ref{orthogonality}). Let $\eta(Y)=Y^m$ for $m=1,2,3...M$. Plugging the best linear estimator $h(Y)=\frac{\gamma}{\gamma+1} Y$ and replacing $Y$ with $X+Z$, we obtain the condition
\begin{equation}
\mathbb{E}\left \{ \left [X-\frac{\gamma}{\gamma+1}(X+Z)\right ](X+Z)^m\right \}=0 \text{ for } m=1,..,M
\end{equation}
Applying the binomial expansion
\begin{equation}
 (X+Z)^m= \sum_{i=0}^{m} {m \choose i}  X^i Z^{m-i}
\end{equation}
and rearranging the terms, we obtain $M$ linear equations that recursively relate the $M+1$ moments of $X$, i.e., for $m=1, ..., M$ we have
\begin{equation}
\label{moment2}
\mathbb{E} (X^{m+1})=\gamma  \mathbb{E} (Z^{m+1}) +  \sum_{i=0}^{m-1} A(\gamma, m,i) \mathbb{E} (Z^{i+1})\mathbb{E} (X^{m-i})
\end{equation}

\noindent where, $A(\gamma,m,i)=\gamma{m \choose i} - {m  \choose {i+1}}$.

Note that if $F_Z(\omega)$ is analytic, $Z$ has finite moments of all orders and $\mathbb E\{|Z^k|\} \leq k! \beta^k$, $ \forall k$. From (\ref{moment2}), by induction, we can show that all moments of $X$ exist and are bounded by $\mathbb E\{|X^k|\} \leq k! (\max\{\gamma,1\} \beta)^k$. This condition is sufficient to show that $X$ also has an analytic characteristic function.
\end{IEEEproof}

The following corollary identifies a case in which a matching source does not exist.
\begin{cor}
For  $\gamma \notin \!  \mathbb N $, if $F_Z(\omega)$ is real and analytic  and it is negative somewhere, i.e., $\exists \omega$ such that $F_Z(\omega)<0 $, then a matching source distribution does not exist.
\end{cor}
\begin{IEEEproof} We prove this corollary by contradiction. Let $F_Z(\omega)$ be a valid characteristic function. Let us first assume that a matching source, $X$, exists. Hence, from Corollary \ref{analytic}, it follows that $X$ must have an analytic characteristic function, $F_X(\omega)$. We will show that this leads to a contradiction. Recall the set of moment equations (\ref{moment2}). It follows by induction over the set of moment equations starting from $m=1$ that, if all odd moments of $Z$ are zero, then so are all odd moments of $X$. As the noise is symmetric, it follows from analyticity of $F_X(\omega)$ that the matching source must also be symmetric, since moments of $X$ fully characterize its distribution.

However, if $\gamma \notin \!  \mathbb N$, by (\ref{main}), it follows that $F_X(\omega)$ is not real everywhere, and hence $f_X(\cdot)$ is not symmetric. This contradiction shows that no matching source exists for symmetric noise distributions which are non positive semi-definite when $\gamma \notin \!  \mathbb N$.
\end{IEEEproof}

Let us provide a commonly used example distribution to which the above corollary applies: uniform distribution over $[-a,a]$. In this case, $f_Z(\cdot)$ is symmetric with an analytic characteristic function, but it is not positive semi-definite. The corollary states that, except for natural values of SNR, the optimal estimator is strictly nonlinear for an additive uniform channel. Example 1 illustrates this point with a numerical example.

{\bf Remark}: As an important application, consider high resolution quantization theory. Standard high resolution approximations assume quantization noise independent of (or uncorrelated with) the source \cite{Gershobook}. In practice, such approximations can be made explicit by using a dithered quantizer \cite{gray1993dq} that generates quantization error independent of the source. Then, the quantizer is equivalent to an additive uniform noise channel. The corollary states that, other than for natural values of SNR, a linear decoder (e.g., a Wiener filter at the decoder) is strictly suboptimal for sources encoded at high resolution or by dithered quantization. 

\subsection{Uniqueness of a Matching Source for a Given Noise}
Note that (\ref{main}) may have multiple solutions due to multiplicity of complex roots. The following corollary establishes that for a large set of source (or noise) distributions, the matching noise (or source) is unique.
\begin{cor}
\label{uniqueness}
If $F_Z(\omega)$ (or  $F_X(\omega)$) is analytic, then the matching $F_X(\omega)$ (or  $F_Z(\omega)$)  is unique.
\end{cor}
\begin{IEEEproof}
We prove this corollary from the set of moment equations (\ref{moment2}). Note that every equation introduces a new variable $\mathbb E( X^{m+1})$, $\text{ for }  m=1,..,M $, so each new equation is linearly independent of its predecessors. Let us consider solving these equations recursively, starting from $m=1$. At each $m$, we have one unknown ($ \mathbb{E} (X^{m+1}$)) in a ``linear" equation. Since the number of equations is equal to the number of unknowns for each $m$, and the equations are linear in terms of the unknown, there must exist a unique moment sequence that solves (\ref{moment2}). From Corollary \ref{analytic}, it also follows that $X$ has an analytic characteristic function. Hence, the moment sequence fully characterizes $X$ and the matching source $X$ (if exists) is unique.
\end{IEEEproof}

\section{Implications of the Linearity Conditions}

In this section, we explore some special cases obtained by varying $\gamma$ and utilizing the matching conditions for MSE and $L_p$. We start with a simple but perhaps surprising result.
\begin{theo}
Given a source and noise of equal variance, the $L_p$ optimal estimator is linear {\em if and only if} the noise and source distributions are identical, i.e., $f_X(x)=f_Z(x), \,\,\, \forall x \in \mathbb R$ and in which case, the optimal estimator is $h(Y)=\frac {1}{2}Y$.
\end{theo}

\begin{IEEEproof} For MSE, it is straightforward to see from (\ref{main}) that, at $\gamma=1$, the characteristic functions must be identical. Since the characteristic function uniquely determines the distribution \cite{billingsley2008probability}, $f_X(x)=f_Z(x)$, $ \forall x \in \mathbb R$. In fact, this results applies more generally. This can be observed directly from Theorem \ref{firsttheo} that $ F_Z(\omega)=F_X(\omega)$ satisfies the necessary and sufficient optimality condition, and hence this result also applies to the $L_p$ norm distortion measure.
\end{IEEEproof}

Our next result pertains to the speciality of Gaussian distribution in the context of linearity of optimal estimation. It is well known that linearity of optimal estimation for all SNR levels characterizes the stable family of distributions, which includes Gaussian as the only finite variance member \cite{rao1947note,allen1938theorem,rothschild1947lois, rao1967some, hardin1982linearity}. However, all prior results on characterizing Gaussian density using linearity of optimal estimation consider optimal estimation for  \textit{all} SNR levels, $\gamma \in \mathbb R^+$.

 Let us consider a setup with given source and noise variables which may be scaled to vary the SNR, $\gamma$. Can the optimal estimator be linear at multiple values of $\gamma$? This question is motivated by the practical setting where $\gamma$ is not known in advance or may vary (e.g., in the design stage of a communication system). It is well-known that the Gaussian source-Gaussian noise pair makes the optimal estimator linear at all $\gamma$ levels. Below, we show that this is the only source-channel pair whose optimal estimators are linear at more than one SNR value.

\begin{figure*}[ht]
\subfigure[SNR=0.1]{
\includegraphics[scale=0.41]{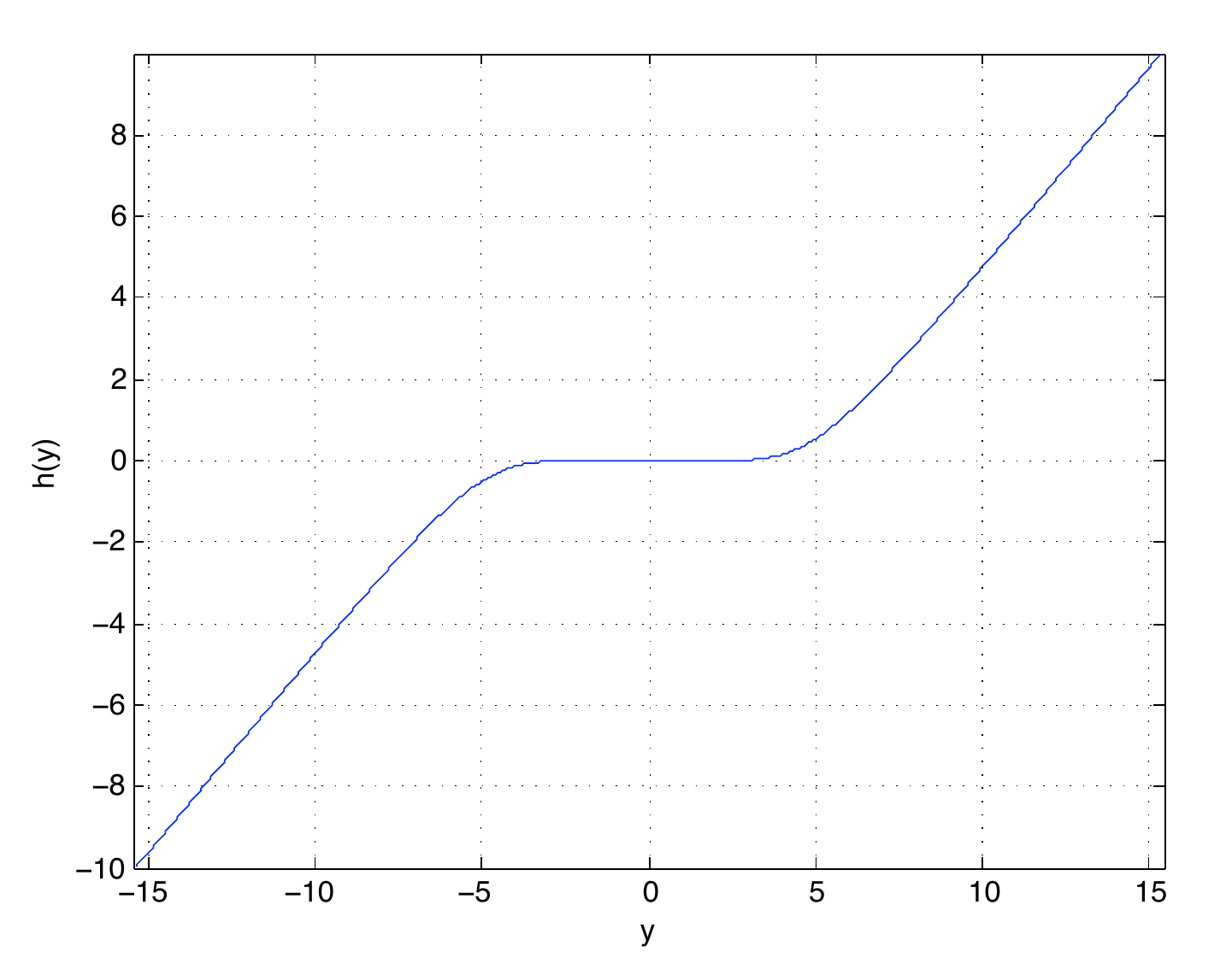}
}
\subfigure[SNR=1]{
\includegraphics[scale=0.41]{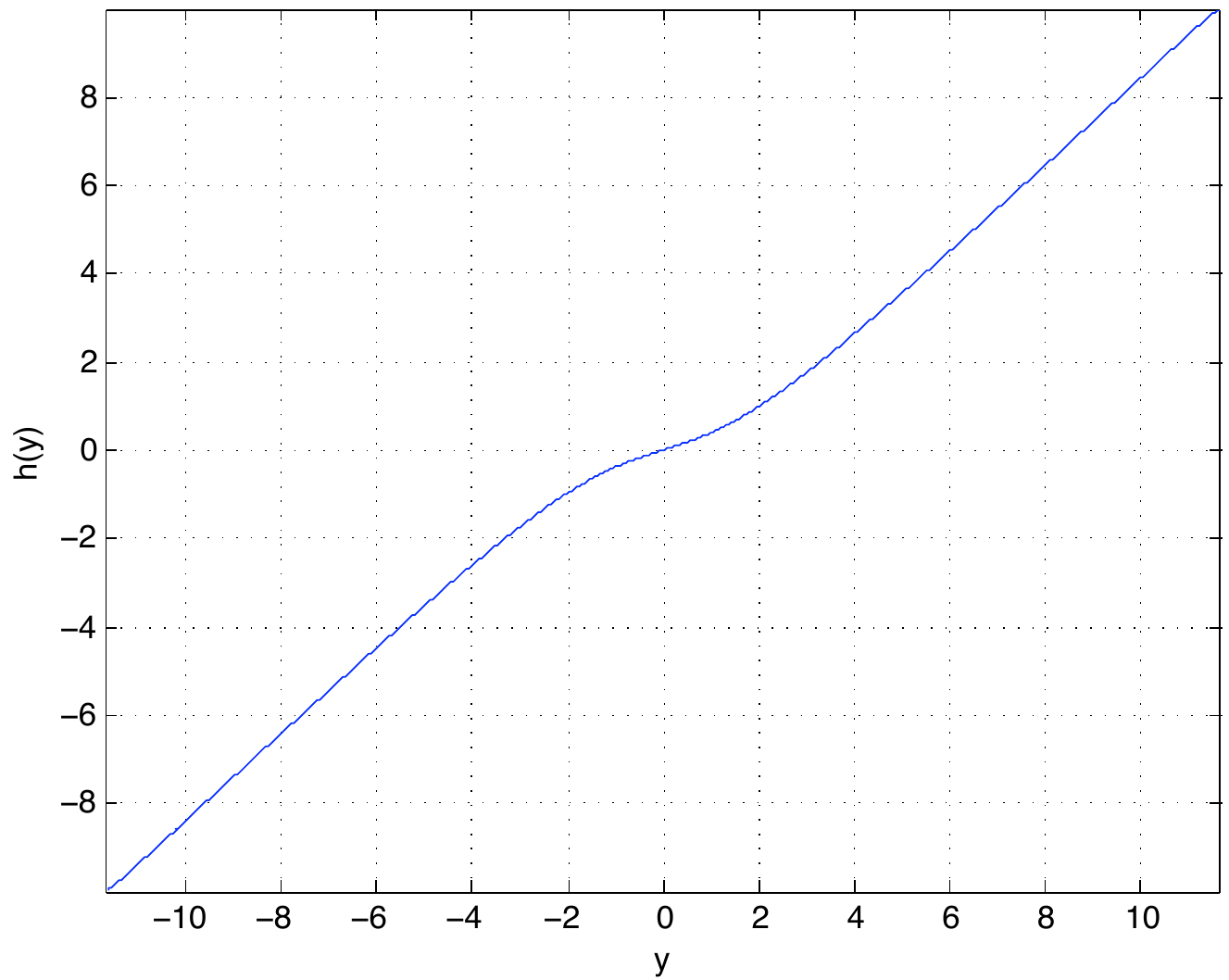}
}
\subfigure[SNR=10]{
\includegraphics[scale=0.41]{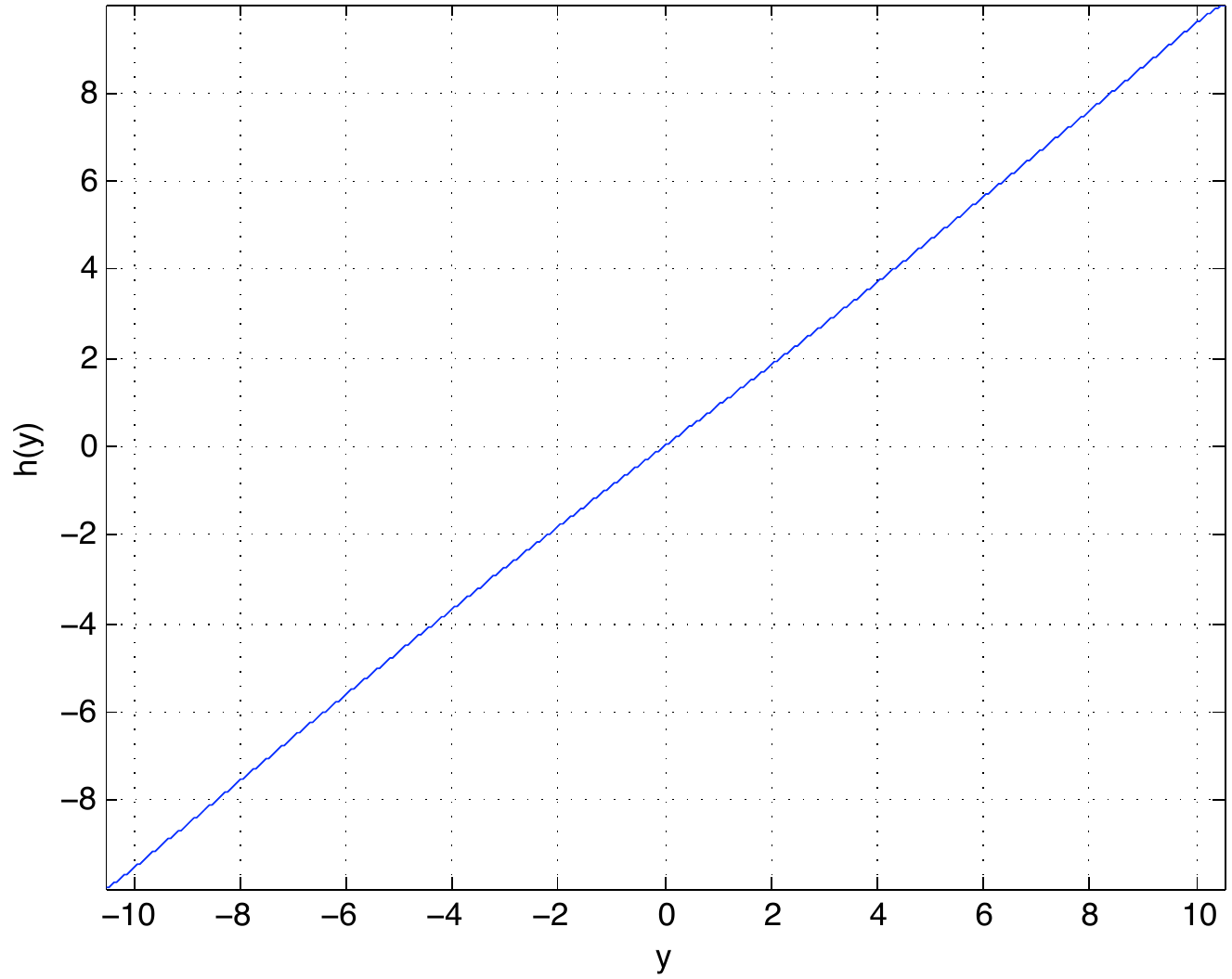}
}
\caption{This figure shows the optimal estimator at various SNR values when $X \sim \mathcal N(0,1)$ and $Z$ is distributed uniformly on the interval $[-a,a]$. The SNR is varied by changing $a$. Observe that the optimal estimator converges to linear as SNR increases.}
\end{figure*}
\begin{theo}
\label{gauss}
Let the source or channel variables be scaled to vary the SNR, $\gamma$. The MSE optimal estimator is linear at two different SNR values $\gamma_1$ and $\gamma_2$, {\em if and only if} source and noise are both Gaussian. Moreover, this claim also holds for $L_p$ norm if the source (or noise) has an analytic characteristic function.\end{theo}

\begin{IEEEproof}
Let $Z_1$ and $Z_2$ denote the noise random variables with variances $\sigma_{z_{1}}^2, \sigma_{z_{2}}^2$ and characteristic functions $F_{Z_1}(\omega),F_{Z_2}(\omega)$ respectively. Let us say the noise is scaled by $\alpha \in \mathbb R$, i.e., $Z_2=\alpha Z_1$ and hence $F_{Z_{2}}(\omega)=F_{Z_1}(\omega\alpha)$ and $\sigma_{z_{2}}^2=\alpha^2\sigma_{z_1}^2 $. Let,
\begin{equation}
\label{snreq}
\gamma_1=\frac{\sigma_x^2}{\sigma^2_{z_1}}, \, \,
\gamma_2=\frac{\sigma_x^2}{\alpha^2 \sigma^2_{z_1}}
\end{equation}
Using (\ref{main}),
\begin{equation}
\label{simple}
F_{X}(\omega)=F_{Z_1}^{\gamma_1}(\omega), F_{X}(\omega)= F_{Z_1}^{\gamma_2}(\omega \alpha)
\end{equation}
Hence,
\begin{equation}
\label{simple2}
F_{Z_1}^{\gamma_1}(\omega) = F_{Z_1}^{\gamma_2}(\omega \alpha)
\end{equation}
\noindent Taking the logarithm on both sides of (\ref{simple2}), applying (\ref{snreq}) and rearranging terms, we obtain
\begin{equation}
\label{eq_up}
\alpha^2=\frac{\log F_{Z_1}( \alpha \omega)}{\log F_{Z_1}(\omega)}
\end{equation}
\noindent  Note that (\ref {eq_up}) should be satisfied for both $\alpha$ and $-\alpha$ since they yield the same $\gamma$.  Hence, $ F_{Z_1}( \alpha \omega)= F_{Z_1}( - \alpha \omega)$ for all $\alpha \in \mathbb R$, which implies $ F_{Z_1}(\omega)= F_{Z_1}( - \omega), \text{ }\forall \omega \in \mathbb R$. Using the fact that the characteristic function is conjugate symmetric (i.e., $F_{Z_1}(-\omega)=F_{Z_1}^*(\omega)$), we get $F_{Z_1}(\omega) \in \mathbb R, \, \forall \omega$.
As $\log F_{Z_1}(\omega)$ is a function from $ \mathbb R  \rightarrow \mathbb C$, Weierstrass theorem \cite{dudley2002real} guarantees that there is a sequence of polynomials that uniformly converges to it:
 $\log F_{Z_1}(\omega)=\sum_{i=0}^{\infty}  k_i \omega^i $, where $k_i \in \mathbb C$. Hence, by (\ref{eq_up}) we obtain:
\begin{equation}
\label{above_eq_last}
\alpha^2=\frac{\sum \limits_{i=0}^{\infty}  k_i (\omega \alpha)^i}{\sum \limits_{i=0}^{\infty}  k_i \omega^i}, \quad  \forall \omega \in \mathbb R,
\end{equation}
which is satisfied for all $\omega$ only if all coefficients $k_i$  vanish, except for $k_2$, i.e., $\log F_{Z_1}(\omega)=k_2\omega^2$, or $\log F_{Z_1}(\omega)=0 \quad \forall \omega \in \mathbb R$ (the solution $\alpha=1$ is of no interest). The latter is not a characteristic function, and the former is the Gaussian characteristic function, $F_{Z_1}(\omega)=e^{k_2\omega^2}$, where we use the established fact that $F_{Z_1}(\omega) \in \mathbb R$.
Since a characteristic function determines the distribution uniquely, the Gaussian source and noise must be the only such pair.

Next, we extend the result to the $L_p$ norm, albeit we require analyticity of the characteristic function of $X$ (or $Z_1$ and $Z_2$). Then, due to Corollary \ref{analytic}, matching noises $Z_1$ and $Z_2$ also have analytic characteristic functions and hence the moments of $X, Z_1$ and $Z_2$ are finite (they have moments of all orders) and moments fully characterize the distribution. The extension to $L_p$ requires a different approach. For simplicity, we first derive the result for MSE (now with analyticity imposed) and then extend the arguments to the $L_p$ case. The following relation between the moments of the original and scaled noise should be satisfied:
\begin{equation}
\label{moment1}
\mathbb{E} (Z_2^{m})= \alpha^m \mathbb{E} (Z_1^{m}) \text{ for } m=1,.., M+1
\end{equation}
\noindent Also, a set of moment equations should hold for two SNR values, $\gamma_1$ and $\gamma_2$. Let us consider the set of moment equations with moments up to $M$:
\begin{equation}
\label{moment3}
\mathbb{E} (X^{m+1}) = \gamma_j  \mathbb{E} (Z_j^{m+1}) + \sum_{i=0}^{m-1} A(\gamma_j,m,i) \mathbb{E} (Z_j^{i+1})\mathbb{E} (X^{m-i})
\end{equation}
 \noindent where  $m=1,..,M, \, j=1,2$ and $A(\gamma,m,i)=\gamma{m \choose i} - {m  \choose {i+1}}$. Similar to the proof of Corollary \ref{uniqueness}, we note that every equation introduces a new variable $\mathbb E( X^{m+1})$, $\text{ for }  m=1,..,M $, so each new equation is independent of its predecessors. Next, we solve these equations recursively, starting from $m=1$. At each $m$, we have three unknowns ($ \mathbb{E} (X^{m+1}),\mathbb{E} (Z_1^{m+1}),\mathbb{E} (Z_2^{m+1})$) that are related ``linearly". Since the number of linearly independent equations is equal to the number of unknowns for each $m$, there must exist a unique solution. We know that the moment sequences of the Gaussian source-channel pair satisfy (\ref{moment3}) since it ensures linearity of optimal estimation. The moment sequence  of a Gaussian satisfies Carleman's general criterion \cite{shohat1943problem} and therefore it uniquely determines the corresponding distribution, so the Gaussian source and noise pair is the only solution to (\ref{moment3}).

 The proof for $L_p$ norm follows the same lines. Note that as mentioned in Sec II.D, the same linear estimator is $L_p$ optimal for a Gaussian source-channel pair. Plugging $Y=X+Z$ in the optimality condition with $L_p$ norm, (\ref{mse5_2}), we reach a similar set of moment equations. Following similar arguments, we show that this result holds for the $L_p$ norm.
 \end{IEEEproof}

 Next, we investigate the asymptotic behavior of optimal estimation at low and high SNR. The results of our asymptotic analysis are of practical importance since they justify the use of linear estimators without recourse to complexity arguments at  high and low asymptotic SNR regimes, under certain conditions. \begin{theo} [for MSE only]
In the limit $\gamma \rightarrow  0 $, the MSE optimal estimator is asymptotically linear if the channel is Gaussian, regardless of the source. Similarly, as $\gamma \rightarrow \infty$, the MSE optimal estimator is asymptotically linear if the source is Gaussian, regardless of the channel.\end{theo}
\begin{IEEEproof} We will present a sketch of the proof here, while a more rigorous formal proof is presented in Appendix C. The proof follows from applying the central limit theorem \cite{billingsley2008probability} to the matching condition (\ref{main}). The central limit theorem states that as $\gamma \rightarrow  \infty $, for any finite variance noise $Z$, the characteristic function of the matching source $F_Z^{\gamma}(\omega)$ pointwise converges to the Gaussian characteristic function. Hence, at asymptotically high SNR, any noise distribution is matched by the Gaussian source.

Similarly, as  $\gamma \rightarrow  0 $ and for any $F_X(\omega)$, $F_X^{\frac{1}{\gamma} }(\omega)$  converges pointwise to the Gaussian characteristic function and hence the MSE optimal estimator is asymptotically linear if the channel is Gaussian.

\end{IEEEproof}

\begin{figure}
\centering
\includegraphics[scale=0.6]{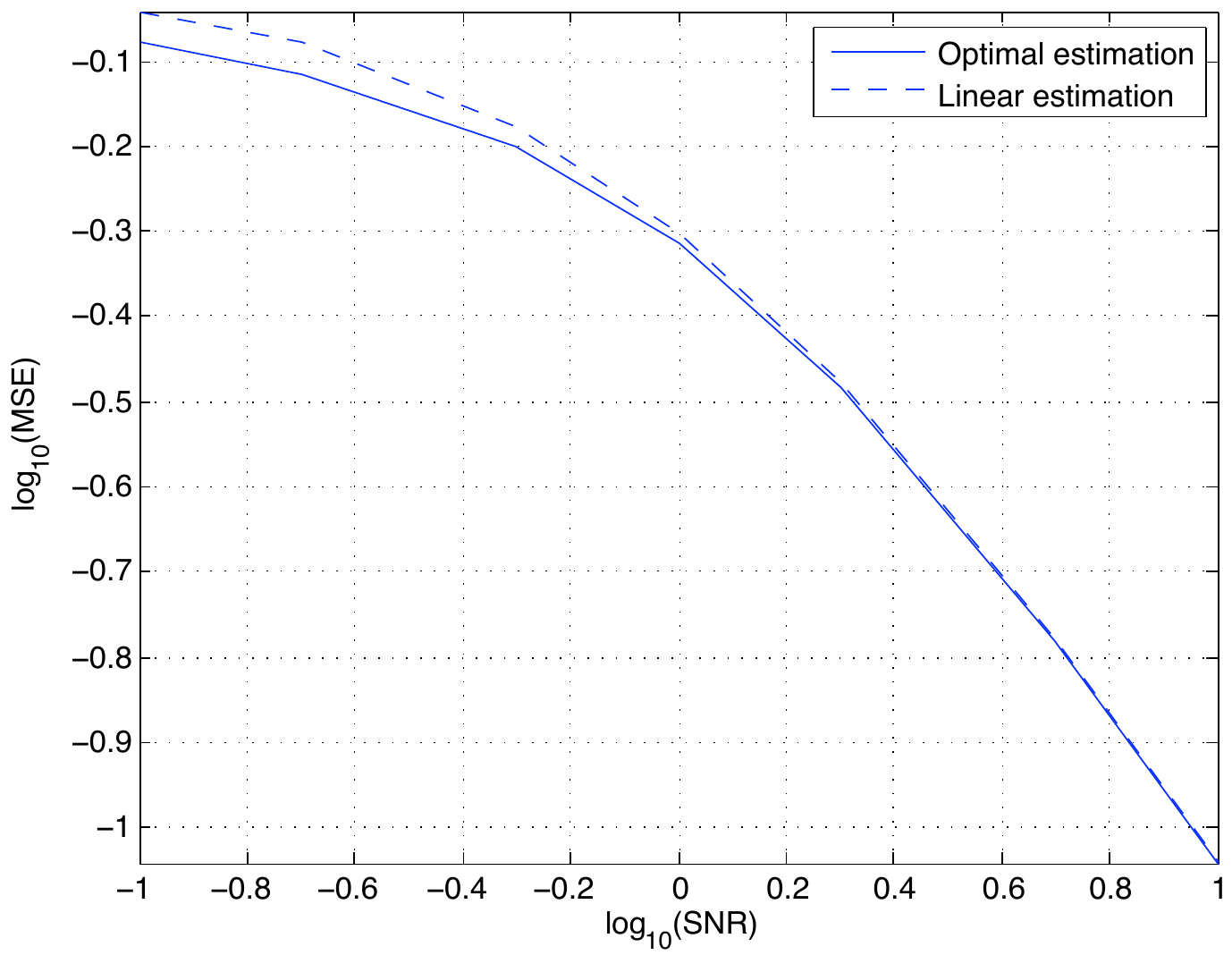}
\label{fig:subfig1}
\caption{This figure shows the variation of estimation error with the channel SNR when $X \sim \mathcal N(0,1)$ and $ Z$ is distributed uniformly on the interval $[-a,a]$. We observe that the error is significant at $\gamma=0.1$ and vanishes at high SNRs.}
\end{figure}

{\bf Example 1}: Let us consider a numerical example that illustrates our findings. Consider a setting where $X$ is Gaussian with unit variance, i.e., $X \sim \mathcal N(0,1)$ and $Z$ is distributed uniformly on the interval $[-a,a]$. Note that this is a typical setting for high rate or dithered quantization of a Gaussian source, in the sense that the  quantization error is uniform and independent of the source. We change $\gamma$ (SNR) by varying $a$ and observe how the optimal estimator ($h(Y)=\mathbb E\{X|Y\}$) and associated estimation error ($\mathbb E\{(X-h(Y))^2\}$) behaves for different $\gamma$.  We numerically calculated the optimal estimator and the estimation error by discretizing the integrals  on a uniform grid, with a step size $\Delta= 0.01$, i.e., to obtain the numerical results, we approximated the integrals as Riemann sums. Figure 2 shows how the optimal estimator  converges to linear as SNR increases. Note that at $\gamma=0.1$, optimal estimator is highly nonlinear while at $\gamma=10$, it practically converges to a linear one. Figure 3 demonstrates how the estimation error varies with SNR. As theoretically expected (and from Figure 2), we see a significant difference at $\gamma=0.1$, while difference vanishes at high SNRs.

 \section{ Extension to Vector Spaces}
Extension of the conditions to the vector case is nontrivial due to the dependencies across components of the source and noise. In this section, for simplicity, we restrict ourselves to the MSE distortion measure. We first give the formal definition of the problem:

We consider the problem of estimating the vector source ${\boldsymbol X} \in \mathbb R^m$ given the observation ${\boldsymbol Y}={\boldsymbol X}+{\boldsymbol Z}$, where $\boldsymbol X$ and ${\boldsymbol Z} \in \mathbb R^m$ are independent, as shown in Figure 1. Without loss of generality, we assume that $\boldsymbol X$ and $\boldsymbol Z$ are zero mean random variables with $m$-fold distributions $f_X(\cdot)$ and $f_Z(\cdot)$. Their respective characteristic functions are denoted $F_X(\boldsymbol \omega)$ and $F_Z(\boldsymbol \omega)$. ${  R_X}=\mathbb E{\{\boldsymbol X \boldsymbol X^T\}}$, ${ R_Z}= \mathbb E\{{\boldsymbol Z \boldsymbol Z^T}\}$ are the covariance matrices of $\boldsymbol X$ and $\boldsymbol Z$, respectively. Let $Q$ be the eigenmatrix of $ R_X  R_Z^{-1}$, and $ U= Q^{-1}$ and let eigenvalues $\lambda_1, ..., \lambda_m$ be the elements of the diagonal matrix $ \Lambda$, i.e., the following holds:
\begin{equation}
\label{eigen}
 R_X  R_Z^{-1}= U^{-1}  \Lambda  U
\end{equation}

We are looking for the conditions on $F_X( \boldsymbol \omega)$ and $F_Z( \boldsymbol \omega)$ such that $ {\boldsymbol h}({\boldsymbol Y}) = K{\boldsymbol Y}$ with $ K=  R_X ( R_X+  R_Z)^{-1} $ minimizes the estimation error $\mathbb E\{ ||{\boldsymbol X-\boldsymbol h(\boldsymbol Y)}||_2^2\}$.

By following a similar approach (details are in Appendix \ref{appc}) to the scalar case we obtain the necessary and sufficient condition of optimality:

\begin{equation}
\label{eq9}
{ U} \nabla \log {F_X({\boldsymbol \omega})}=  \Lambda{ U} \nabla \log {F_Z({\boldsymbol \omega})}
\end{equation}

We will make use of the following auxiliary lemma from matrix analysis.
\begin{lemma}
\label{ax}
Given a function $f:\mathbb R^n \rightarrow \mathbb R$, matrix ${ A} \in \mathbb R ^{n\times m}$ and vector ${\boldsymbol x} \in \mathbb R^m$
\begin{equation}
\nabla_x f ({ A \boldsymbol x})={ A}^T \nabla f({ A \boldsymbol x})\end{equation}
\end{lemma}
\begin{IEEEproof}
See Appendix \ref{appd}.
\end{IEEEproof}

Next, we state the main theorem in vector settings.

 \begin{theo}
 \label{vectortheo}
 Let the characteristic functions of the transformed source and noise ($ U \boldsymbol  X$ and $ U\boldsymbol Z$) be $F_{UX}({\boldsymbol \omega})$ and $F_{UZ}({\boldsymbol \omega})$. The necessary and sufficient condition for linearity of optimal estimation is:
\begin{equation}
\label{abeq2}
\frac {\partial   \log {F_{UX}({\boldsymbol \omega})}}{\partial \omega_i}= \lambda_i \frac {\partial  \log {F_{UZ}({\boldsymbol \omega})} }{\partial \omega_i}, 1\leq i \leq m
\end{equation}
 \end{theo}

\begin{IEEEproof}
Let us define $\boldsymbol  {\tilde   \omega} =  (U^{-T})\boldsymbol \omega$, hence $\boldsymbol  {  \omega} =  U^{T}\boldsymbol{ \tilde \omega}$. Plugging this in  (\ref{eq9}), we have
\begin{equation}
\label{up10}
{U} \nabla_{ U^{T}\boldsymbol{ \tilde \omega}} \log F_X( U^{T}\boldsymbol{ \tilde \omega})=  \Lambda{  U} \nabla_{ U^{T}\boldsymbol{ \tilde \omega}} \log F_Z(  U^{T} \boldsymbol{ \tilde \omega})
\end{equation}
Using Lemma \ref{ax}, we can rewrite (\ref{up10}) as
\begin{equation}
\label{vector4}
 \nabla_{\boldsymbol { \tilde \omega}} \log F_X( U^{T}\boldsymbol{ \tilde \omega})=  \Lambda \nabla_{\boldsymbol{ \tilde \omega}} \log F_Z(  U^{T}\boldsymbol{ \tilde \omega})
\end{equation}
Note that the characteristic functions of the source and noise after transformation can be written in terms of the known characteristic functions  $F_X(\boldsymbol \omega)$ and $F_Z(\boldsymbol \omega)$, specifically $ F_{UX}({\boldsymbol \omega})= {F_X({ U^{T}} {\boldsymbol \omega})} $ and $ F_{UZ}({\boldsymbol \omega})=  {F_Z(  U^{T} {\boldsymbol \omega})}$.
Plugging these expressions in (\ref{vector4}), we have
\begin{equation}
\label{up11}
 \nabla_{\boldsymbol { \tilde \omega}} \log F_{UX}(\boldsymbol{ \tilde \omega})=  \Lambda \nabla_{\boldsymbol{ \tilde \omega}} \log F_{UZ}( \boldsymbol{ \tilde \omega})
\end{equation}
Using the fact that $ \Lambda $ is diagonal, we convert (\ref{up11}) to the set of $m$ scalar differential equations of (\ref{abeq2}).

\end{IEEEproof} 

Further insight into the above necessary and sufficient condition is provided via the following corollaries.
\begin{cor}
\label{corr2}
Let  $F_{[UX]_i}(\omega)$ and $F_{[UZ]_i}(\omega)$ be the marginal characteristic functions of the transform coefficients $[U\boldsymbol X]_i$ and $[U\boldsymbol Z]_i $ respectively. A necessary condition for linearity of optimal estimation is: 
\begin{equation}
\label{abeq}
 F_{[UX]_i}(\omega)=F_{[UZ]_i} ^{\lambda_i}(\omega), 1 \leq i\leq m
\end{equation}
\end{cor}

\begin{IEEEproof} The marginal characteristic functions of $[U \boldsymbol X]_i$ and $[U \boldsymbol Z]_i$ are obtained by setting $\omega_k = 0$,  $\forall k\neq i$ in  $F_{UX}(\boldsymbol{ \omega})$ and $F_{UZ}(\boldsymbol{ \omega})$ respectively. By setting $\omega_k = 0$,  $\forall k\neq i$ in both sides of  (\ref{abeq2}), we have 
\begin{equation}
\label{abeq3}
\frac {\partial   \log {F_{[UX]_i}({\omega})} }{\partial \omega}= \lambda_i \frac {\partial   \log {F_{[UZ]_i}(\omega)} }{\partial \omega}, \text{  } 1 \leq i\leq m
\end{equation}
The solution to this differential equation is given by:
\begin{equation}
\log{ F_{[UX]_i}(\omega)}= \lambda_i \log{ F_{[UZ]_i}(\omega)}+C
\end{equation}
where $C$ is a constant. Imposing $F_{[UZ]_i}(0)=F_{[UX]_i}(0)=1$, we obtain $C=0$, which implies:
\begin{equation}
\label{mainvector}
{F_{[UX]_i}(\omega)}= F_{[UZ]_i}^{\lambda_i}(\omega),1\leq i \leq m
 \end{equation}  
\end{IEEEproof}

\begin{cor}
A necessary condition for linearity of optimal estimation is that one of the following holds for every pair  $i, j $, $1 \leq i,j \leq m$: 
\begin{itemize}
\item i) ${ \lambda}_i= { \lambda}_j$
\item ii) $[U \boldsymbol X]_i$ is independent of $ [U \boldsymbol X]_j$ and $[ U\boldsymbol Z]_i$ is independent of $[U  \boldsymbol Z]_j$. 
 \end{itemize}
\end{cor}

\begin{IEEEproof}
Let us rewrite (\ref{abeq2}) explicitly for the $i^{th}$ and $j^{th}$ coefficients.  
  \begin{equation}
\label{vectori}
 \frac {\partial   \log {F_{UX}({\boldsymbol \omega})} }{\partial \omega_i}= \lambda_i \frac {\partial   \log {F_{UZ}( {\boldsymbol \omega})} }{\partial \omega_i}
\end{equation}
 \begin{equation}
\label{vectorj}
 \frac {\partial   \log {F_{UX}({ \boldsymbol \omega})} }{\partial \omega_j}= \lambda_j \frac {\partial   \log {F_{UZ}( {\boldsymbol \omega})} }{\partial \omega_j}
\end{equation}
The partial derivative of both sides of  (\ref{vectori})  with respect to $\omega_j$ and both sides of  (\ref{vectorj}) with respect to $\omega_i$, to obtain the following:
 \begin{equation}
\label{vectorii}
 \frac {\partial^2   \log {F_{UX}( {\boldsymbol \omega})} }{\partial \omega_i\partial \omega_j}= \lambda_i \frac {\partial^2   \log {F_{UZ}({\boldsymbol \omega})} }{\partial \omega_i \partial \omega_j}
\end{equation}
\begin{equation}
\label{vectorjj}
 \frac {\partial^2   \log {F_{UX}({\boldsymbol \omega})} }{\partial \omega_i \partial \omega_j}= \lambda_j \frac {\partial^2   \log {F_{UZ}({\boldsymbol \omega})} }{\partial \omega_i\partial \omega_j}
\end{equation}
There are only two ways to simultaneously satisfy (\ref{vectorii}) and (\ref{vectorjj}): i) $\lambda_i=\lambda_j$ ii) the second order derivatives vanish, i.e., 
\begin{equation}
\label{upeqr}
\frac {\partial^2   \log F_{UX}({\boldsymbol \omega}) }{\partial \omega_i \partial \omega_j}=0
\end{equation}
\begin{equation}
\frac {\partial^2   \log F_{UZ}({\boldsymbol \omega})} {\partial \omega_i \partial \omega_j}=0
\end{equation}

Let us focus on $\boldsymbol X$ i.e., (\ref{upeqr}), derivation for $\boldsymbol Z$ follows similarly. $F_{[UX]_{ij}}( \omega_i, \omega_j)$, i.e., the marginal characteristic function of the pair $([U \boldsymbol X]_i,[U \boldsymbol X]_j)$ is obtained by setting $\omega_k = 0$,  $\forall k\neq i,j$. Then,  (\ref{upeqr}) implies 
\begin{equation}
\label{uper}
\frac {\partial^2   \log F_{[UX]_{ij}}({ \omega_i, \omega_j}) }{\partial \omega_i \partial \omega_j}=0
\end{equation}
which means
\begin{equation}
\label{upeqr3}
\log F_{[UX]_{ij}}({ \omega_i, \omega_j}) =A(\omega_i) +B(\omega_j) 
\end{equation}
for some functions $A$ and $B$, i.e., $\log F_{[UX]_{ij}}({ \omega_i, \omega_j})$ is additively separable in terms of $\omega_i$ and $\omega_j$. This implies
\begin{equation}
\label{upeqr4}
F_{[UX]_{ij}}({ \omega_i, \omega_j}) =C(\omega_i) D(\omega_j) 
\end{equation}
for some functions $C$ and $D$. But (\ref{upeqr4}) implies independence of the $i^{th}$ and $j^{th}$ transform coefficients of source $\boldsymbol X$. The independence of the $i^{th}$ and $j^{th}$ transform coefficients of the noise $\boldsymbol Z$ follows from similar arguments.
\end{IEEEproof}

\begin{cor}
\label{corr3}
If the necessary condition of Corollary \ref{corr2} is satisfied, then a sufficient condition for linearity of optimal estimation is that $ U$ generates independent coefficients for both $\boldsymbol X$ and $\boldsymbol Z$. 
\end{cor}
\begin{IEEEproof}
Independence of the transform coefficients implies that the joint characteristic function is the product of the marginals:
\begin{equation}
\label{abeq9}
F_{UX}({\boldsymbol \omega})=\prod_{i=1}^m F_{[UX]_i}(w_i), \text{      } F_{UZ}({\boldsymbol \omega})=\prod_{i=1}^m F_{[UZ]_i}(w_i)
\end{equation}
Plugging (\ref{abeq9}) into the necessary and sufficient condition (\ref{abeq2}) of Theorem \ref{vectortheo}, it is straightforward to show that (\ref{abeq}), the necessary condition of Corollary \ref{corr2}, is now both necessary and sufficient.
\end{IEEEproof}

While the condition in Corollary \ref{corr3} involves independence of transform coefficients, the weaker property of uncorrelatedness is already guaranteed by transform $ U$. The matrix $ U$ diagonalizes both $ R_X$ and $ R_Z$. We formalize this in the following lemma:
\begin{lemma}
Transform $ U$ decorrelates both source and noise: both $ U  R_X  U^{T}$ and $ U  R_Z  U^{T}$ are diagonal matrices.
\end{lemma}

\begin{IEEEproof}
Since both $ R_X$ and $ R_Z$ are, by definition, positive definite matrices, there exists a matrix $ S$ that simultaneously diagonalizes $ R_X$ and whitens $ R_Z$, i.e., $  S  R_X  S^T= \Lambda_X$ and $ S  R_Z  S^T= I$ where $ \Lambda_X$ is diagonal and $ I$ is the identity matrix \cite{hornma}. Hence, $ R_X$  and $ R_Z$ can be expressed as the following:
\begin{equation}
\label{e1}
 R_X =   S^{-1} \Lambda_X  S ^{-T}  , \text{      }      R_Z=   S^{-1}   S^{-T}
\end{equation}
Plugging (\ref{e1}) into (\ref{eigen}) we obtain $ U=  \Lambda_U  S$, where $ \Lambda_U$ is diagonal. Substituting $ U$ in $ U   R_X  U^{T}$ and $ U   R_Z  U^{T}$, we obtain:
\begin{equation}
 U  R_X  U^{T}= \Lambda_U  \Lambda_X  \Lambda_U^T  , \text{      }   U  R_Z  U^{T}= \Lambda_U  \Lambda_U^T
\end{equation}
 The product of diagonal matrices is also diagonal.
\end{IEEEproof}

As an example where the optimal estimator is known to be linear, consider the multivariate Gaussian case. Note that the Gaussian source-channel pair satisfies the scalar matching condition for any SNR, i.e., (\ref{mainvector}). As any linear transform preserves joint Gaussianity in the transform domain, $U$ generates jointly Gaussian and uncorrelated coefficients which are therefore independent, satisfying the conditions of Corollary \ref{corr3}.

Another, perhaps surprising, example where the optimal estimator is linear involves identically distributed source $\boldsymbol X$ and noise $\boldsymbol Z$. In this case, the linear estimator is optimal {\em irrespective of the distribution} of source and noise. It is straightforward to show that the necessary and sufficient conditions of Theorem 6 are satisfied if $F_X(\boldsymbol \omega)=F_Z(\boldsymbol \omega)$.

{\bf Example 2}: Let us consider a numerical example that highlights the differences in conditions derived for vectors from the scalars. Consider a setting where a two dimensional random variable  $\boldsymbol Z'$ has independent components, both of which are uniformly distributed over $[-a, a]$, i.e., $\boldsymbol Z'=[Z_1', Z_2']$ and $Z_1'\sim Z_2' \sim U[-a,a]$. Also, let $\boldsymbol X'$ have two independent identically distributed components $\boldsymbol X'=[X_1', X_2']$ where $X_1'$ and $X_2'$ are distributed according to a density given by the convolution of the uniform density with itself, i.e., ${X_1'}\sim {X_2'}  \sim \left( U[-a,a] \ast U[-a,a] \right)$. Since $\boldsymbol X'$ and $\boldsymbol Z'$ satisfy the sufficient conditions in Corollary 8, the optimal estimator is linear for the source-channel pair $(\boldsymbol X', \boldsymbol Z' )$. \\

Let us next consider the source-channel pair $(\boldsymbol X,\boldsymbol Z)$ to be $\boldsymbol X= Q_X \boldsymbol X'$ and  $\boldsymbol Z= Q_Z \boldsymbol Z'$ where $ Q_X$ and $  Q_Z$ are  $2 \times 2$ orthogonal matrices ($ Q_X   Q_X^T=  Q_Z   Q_Z^T= I$). This introduces dependencies among the components of $\boldsymbol X$ and $\boldsymbol Z$. We already saw that for $ Q_X =  Q_Z =  I$, the optimal estimator is linear. Also, from standard linear estimation principles \cite{kailath_book}, it follows that the minimum estimation error achievable by linear estimators does not depend on $ Q_X$ and $ Q_Z$, i.e., linear estimation error is a constant with respect to $ Q_X$ and $ Q_Z$. The question we are interested in is - can the linear estimator be optimal for any other pair $( Q_X, Q_Z)$? Corollary \ref{corr3} sheds light on this question. First, we consider the case where $ Q_X=\pm Q_Z$. Observe that, any orthogonal matrix $ U$ satisfies condition (\ref{eigen}). Hence, we can set $ U= Q_X^{-1}=\pm Q_Z^{-1}$ leading to $ U \boldsymbol X=\boldsymbol X'$ and $ U \boldsymbol Z=\boldsymbol Z'$. This implies that $ U \boldsymbol X$ and $ U \boldsymbol Z$ satisfy conditions in Corollary 8, which are sufficient to prove linearity of optimal estimators. Hence, for the source-channel pair $(\boldsymbol X,\boldsymbol Z)$, optimal estimators are always linear if $ Q_X=\pm Q_Z$.

Finally, we consider the case where $ Q_X \neq \pm Q_Z$. In general, any orthogonal matrix can be written in terms of another orthogonal matrix as
\begin{equation}
 Q_X= G(\theta)  Q_Z
\end{equation}
where $ G(\theta)=\left[ \begin{array}{ccc}
\cos(\theta) & -\sin(\theta)  \\
\sin(\theta) & \cos(\theta)  \end{array} \right]  $ (also known as  Givens rotation \cite{hornma}). For a constant $ Q_Z $, we change $ Q_X $ by varying $\theta$ and observe the behavior of the difference between the mean square errors obtained by the optimal and the linear estimators. As a performance metric, we consider the normalized difference of estimation errors, i.e., (MSE of linear estimation-MSE of optimal estimation)/ MSE of optimal estimation. The variation of the normalized difference as a function of $\theta$ is plotted in Figure 4. Observe that, at $\theta=0$ and $\pi$ the optimal estimator is linear as expected from Corollary \ref{corr3}. It is not hard to show using symmetry of $\boldsymbol X'$ and $\boldsymbol Z'$ that the conditions of Corollary \ref{corr3} are also satisfied for $\theta=\pi/2$ (and $3\pi/2$). A perhaps interesting observation is that the deviation of optimal estimator from linearity grows monotonically in $\theta$ in the range $\theta \in (0,\pi/4)$.

\begin{figure}
\centering
\includegraphics[scale=0.6]{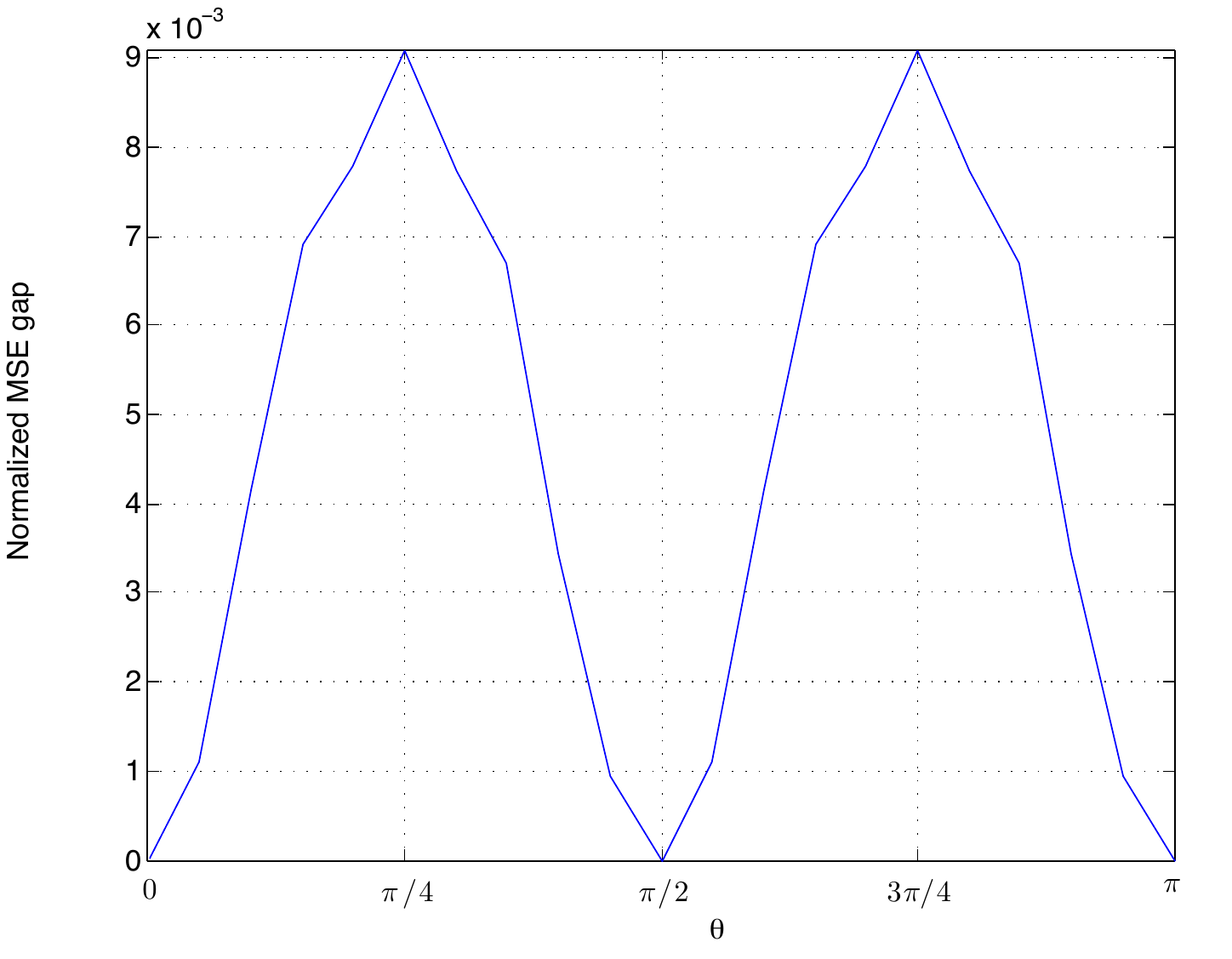}
\label{figure4}
\caption{Normalized difference between optimal and linear estimation versus the Givens rotation parameter $\theta$, for the source channel pair $(\boldsymbol X, \boldsymbol Z)$.}
\end{figure}

An important observation is that the necessary and sufficient condition for scalars (\ref{main}) is also a necessary condition for vectors (\ref{abeq}), in the transform domain. Due to this fact, it is straightforward to extend the existence and uniqueness results and implications of the scalar matching conditions to the vector spaces. These trivial extensions are omitted here for conciseness. 

\section{Conclusion}
In this paper, we derived conditions under which the $L_p$ optimal estimator is linear. We identified the conditions for the existence and uniqueness of a source distribution that matches the noise in a way that ensures linearity of the optimal estimator, for the special case of $p=2$. One trivial example of this type of matching occurs for Gaussian source and Gaussian noise at all SNR levels. Another instance of matching happens when the source and noise are identically distributed. We also showed that the Gaussian source-channel pair is unique in that it is the only pair for which the optimal estimator is linear at more than one SNR value. Moreover, we showed the asymptotic linearity of MSE optimal estimators at low SNR if the channel is Gaussian, regardless of the source, and vice versa, at high SNR if the source is Gaussian regardless of the channel. We also studied the extension to vector spaces where additional conditions are derived beyond those inherited from the scalar case, which concern interactions across components.

\appendices
\section{ Proof of Lemma \ref{lemma1}}
\label{appa}
\begin{IEEEproof}
First, we show the sufficiency of the necessary conditions for $L_p$ norm. Note that $\Phi(x)=|x|^p$  is convex for $p\geq 2$, i.e., $\frac{d^2 |x|^p}{dx^2}\geq 0$,  $\forall x-\{0\}$. We need to show $\frac{\partial^2}{\partial^2 \epsilon}  J\left [h(y)+\epsilon \eta( y)\right ] \bigg|_{\epsilon =0} \geq0$, for any $\eta(y)$ variation function.
\begin{align}
\frac{\partial^2}{\partial^2 \epsilon} J&\left [h(y)+\epsilon \eta( y)\right ] \bigg|_{\epsilon =0}= \nonumber \\
 &\int \int \eta^2 (y) \Phi^{''}(x-h(y)) f_X(x) f_{Z}(y-x)dx dy
\end{align}
All factors in the integral are non-negative and hence, $\frac{\partial^2}{\partial^2 \epsilon}  J\left [h(y)+\epsilon \eta( y)\right ] \bigg|_{\epsilon =0} \geq 0$, for any $\eta(y)$.\\
Next, we show the uniqueness (in probabilistic sense) of the optimal estimator for even natural $p$. Assume $h_1(Y)$ and $h_2(Y)$ both satisfy (\ref{mse5_2}) while $\mathbb P\left [  h_1(Y)\neq h_2(Y)  \right] >0$,  i.e.,  over a set of positive measure $h_1(Y)\neq h_2(Y)$. Then, the following holds for any $\eta(Y)$
\begin{equation}
\label{peven}
\mathbb E \left \{ \{ [X-h_2(Y)]^{p-1}-[X-h_1(Y)]^{p-1}\} \eta(Y)\right \}=0 
\end{equation}
Note that
\begin{equation}
\label{up14}
[X-h_2(Y)]^{p-1}-[X-h_1(Y)]^{p-1} = (h_1(Y)-h_2(Y)) \beta(X,Y)
\end{equation}
where
\begin{equation}
 \beta(X,Y)= \displaystyle \sum_{m=0}^{p-2} [X-h_1(Y)]^{p-2-m} [X-h_2(Y)]^m
\end{equation}
\begin{proposition}
$h_1(Y)\!\neq\!h_2(Y)$ implies $\beta(X,Y)>0$ $\forall X,Y \in \mathbb R$.
\end{proposition}
To see this, we note that (\ref{up14}) is a simple factorization of the form 
\begin{equation}
A^{p-1}-B^{p-1} = (A-B) P (A,B)
\end{equation}
where $P(A,B)$ is a polynomial. Now if $A \neq B$, then the sign of left hand side equals to the sign of $A-B$. Hence $P (A,B)>0$. 

Next, plugging  $\eta(Y)=h_1(Y)-h_2(Y)$ in (\ref{peven}), we obtain,
\begin{equation}
\mathbb E \left \{  [h_1(Y)-h_2(Y) ]^2 \beta(X,Y) \right \}=0 
\label{suff2}
\end{equation}
Since $h_1(Y)\neq h_2(Y)$ implies $\beta(X,Y)>0$  $\forall X, Y \in \mathbb R$, then (\ref{suff2}) requires $h_1(Y)= h_2(Y)$ almost everywhere, contradicting the hypothesis $\mathbb P\left [  h_1(Y)\neq h_2(Y)  \right] >0$.

\end{IEEEproof}

\section{ Proof of Theorem \ref{firsttheo} }
\label{appb}
\noindent The necessary and sufficient condition  (\ref{mse5_2}) can be rewritten as: 
 \begin{equation}
\int \left \{ \int  (x-ky)^{p-1} f_X(x)f_Z(y-x) dx \right \} \eta(y)  dy =0
\end{equation}
for all admissible perturbation functions $\eta(y)$. This equality is achieved  for all $\eta(y)$ if and only if  the  
expression in braces vanishes  almost everywhere. Hence, (\ref{mse5_2}) is satisfied if and only if:
 \begin{equation}
 \label{first}
\Psi(y)\triangleq\int (x-ky)^{p-1} f_X(x)f_Z(y-x)dx =0, a.e.
\end{equation}
Applying the binomial expansion to the first factor
\begin{equation}
(x-ky)^{p-1}= \sum_{m=0}^{p-1} {p-1 \choose m}  (-ky)^m x^{p-m-1}
\end{equation}
and rearranging terms, we get
 \begin{equation}
\sum_{m=0}^{p-1}   \binom {p-1}{m} (-ky)^m \int x^{p-1-m} f_X(x)f_Z(y-x)dx=0
\label{above_eq}
\end{equation}
Let $\ast$ denote the convolution operator, and rewrite (\ref{above_eq}) as
\begin{equation}
\label{above_eq2}
\sum_{m=0}^{p-1}   \binom {p-1}{m}  (-ky)^m \left [y ^{p-1-m}f_X(y) \ast f_Z(y)\right]=0
\end{equation}
Taking the Fourier transform\footnote{Note that the Fourier transforms exist due to the finite moments assumption stated in Section II.A.},
\begin{equation}
\displaystyle \sum_{m=0}^{p-1}   \binom {p-1}{m}  (-k)^m \frac{d^m}{d\omega^m}  \left [\frac{d^{p-1-m}(F_X(\omega))}{d \omega ^{p-1-m}}   F_Z(\omega)\right]=0
\end{equation}
differentiating in parts,
\begin{equation}
\displaystyle \sum_{m=0}^{p-1}   \binom {p-1}{m}  (-k)^m   \sum_{l=0}^{m}   \binom {m}{l}  \frac{d^{p-1-l}F_X(\omega)}{d \omega ^{p-1-l}}    \frac{d^l F_Z(\omega)} {d \omega ^{l}} =0
\end{equation}
interchanging summations,
\begin{equation}
\displaystyle  \sum_{l=0}^{p-1}  \frac{d^{p-1-l}F_X(\omega)}{d \omega ^{p-1-l}}    \frac{d^l F_Z(\omega)} {d \omega ^{l}}  \sum_{m=l}^{p-1}   \binom {p-1}{m}  (-k)^m     \binom {m}{l} =0
\end{equation}
applying some combinatoric algebra,
\begin{align}
&\displaystyle  \sum_{l=0}^{p-1}   \binom {p-1}{l}  \frac{d^{p-1-l}F_X(\omega)}{d \omega ^{p-1-l}} \frac{d^l F_Z(\omega)} {d \omega ^{l}}   \nonumber \\
  & \sum_{m=l}^{p-1}   \frac {(p-1-l)!}   {(m-l)! (p-1-m)! }  (-k)^m =0
\end{align}
and substituting $t=m-l$, we get
\begin{align}
\displaystyle  \sum_{l=0}^{p-1}   \binom {p-1}{l}  &\frac{d^{p-1-l}F_X(\omega)}{d \omega ^{p-1-l}}    \frac{d^l F_Z(\omega)} {d \omega ^{l}}  \nonumber \\
&\sum_{t=0}^{p-1-l}   \binom {p-1-l}   {t}  (-k)^{(t+l)} =0
\end{align}
Finally, noting that
\begin{equation}
\label{last}
  (1-k)^{p-1-l}= \sum_{t=0}^{p-1-l}   \binom {p-1-l}   {t}  (-k)^{t}
  \end{equation}
we obtain that (\ref{main_Lp}) is a necessary and sufficient condition. 

We note that all steps of the derivation were obtained as `` if and only if " statements, hence the converse is automatically proved. 

 \section{Formal Proof of Theorem 5}
Let  $h(y)=k[y+\xi(y)]$ be the polynomial expansion of the optimal estimator where $\xi(y)$ consists of terms with order only two or higher. Let us rewrite the optimal estimator,
\begin{equation}
h(y)=k[y+\xi(y)]= \frac { \int { x}  f_X( { x})   f_Z ({ y- x})\, { dx} }{\int f_X( { x})   f_Z ({ y- x})\, { dx}}
\end{equation}
or 
\begin{equation}
k[y+\xi(y)] \int f_X( { x})   f_Z ({ y- x})\, { dx}= \int { x}  f_X( { x})   f_Z ({ y- x})\, { dx}
\end{equation}
Expressing the integrals as convolutions, we have
\begin{equation}
k[y+\xi(y)] \left [f_X({ y}) \ast f_Z({ y})\right]=\left [ { y} f_X({ y})\right ] \ast f_Z({ y})
\end{equation}
Taking the Fourier transform of both sides, we obtain 
\begin{align}
j { k} \frac { d \left[ F_X({  \omega})F_Z({  \omega})\right ]}{dw}+&k  [ F_X({  \omega})F_Z({  \omega}) ]  \ast  \Xi(\omega) \nonumber \\
& = j F_Z({  \omega}) \frac{d  F_X({  \omega})}{dw}
\end{align}
where $\Xi(\omega)$ denotes the Fourier transform of $\xi(\cdot)$. Plugging  $k=\frac{\gamma}{1+\gamma}$ and dividing both sides by  $F_X({  \omega})F_Z({  \omega})$ we have
\begin{equation}
\frac{1}{F_X(\omega)}  \frac {dF_X(\omega)}{d\omega}=\frac{\gamma}{1+\gamma} \zeta(\omega)+\gamma \frac{1}{F_Z(\omega)}  \frac {dF_Z(\omega)}{d \omega}
\end{equation}
or more compactly,
\begin{equation}
 \frac{ d}{d\omega} \log {F_X(\omega)}= \frac{\gamma}{1+\gamma} \zeta(\omega)+ \frac { d}{d\omega} \log {F_Z^{\gamma}(\omega)}
\end{equation}
where $\zeta(\omega)\triangleq\frac{\gamma}{1+\gamma} \frac{[ F_X({  \omega})F_Z({  \omega}) ]  \ast  \Xi(\omega)}{j [ F_X({  \omega})F_Z({  \omega}) ] }$. 

Now consider the setting where the source is Gaussian and  $\gamma \rightarrow \infty$. By applying the central limit theorem, we have ${F_Z^{\gamma}(\omega) \rightarrow F_X(\omega)}$ pointwise as $\gamma \rightarrow \infty$. Hence, $\zeta(\omega)\rightarrow 0$ pointwise for all $\omega \in \mathbb R$. But this implies $\Xi(\omega)\rightarrow 0$ (pointwise) and hence  in the limit $\gamma\rightarrow \infty$, $\xi(y)=0$ almost everywhere with respect to the density of $y$. Also, it follows from the same arguments that when the noise is Gaussian and $\gamma \rightarrow 0$, $\xi(y)=0$ a.e. 

 \section{Derivation-Vector Case}
 \label{appc}
 \noindent Let us rewrite the MSE optimal estimator for the vector case:
  \begin{equation}
  \label{veceq}
\boldsymbol h({\boldsymbol y}) = \frac { \int {\boldsymbol x}  f_X( {\boldsymbol x})   f_Z ({\boldsymbol y-\boldsymbol x})\, {\boldsymbol dx} }{\int f_X( {\boldsymbol x})   f_Z ({\boldsymbol y-\boldsymbol x})\, {\boldsymbol dx}}
\end{equation}
 Plugging $\boldsymbol h({\boldsymbol  y})= K \boldsymbol y$ in (\ref{veceq}) we obtain,
 \begin{equation}
 { K} {\boldsymbol y} \int f_X( {\boldsymbol x})   f_Z ({\boldsymbol y-\boldsymbol x})\, {\boldsymbol dx}= \int {\boldsymbol x}  f_X( {\boldsymbol x})   f_Z ({\boldsymbol y-\boldsymbol x})\, {\boldsymbol dx}
\end{equation}
Expressing the integrals as $m$-fold convolutions, we get
\begin{equation}
{ K}  {\boldsymbol y} \left [f_X({\boldsymbol y}) \ast f_Z({\boldsymbol y})\right]=\left [ {\boldsymbol y} f_X({\boldsymbol y})\right ] \ast f_Z({\boldsymbol y})
\end{equation}
Taking the Fourier transform of both sides,
\begin{equation}
j { K} \nabla \left [ F_X({\boldsymbol  \omega})F_Z({\boldsymbol  \omega})\right ]  =j F_Z({\boldsymbol  \omega}) \nabla F_X({\boldsymbol  \omega})
\end{equation}
and rearranging terms, we get
\begin{equation}
\left ({ I- K }\right ) \frac{1}{F_X({\boldsymbol \omega})}  \nabla {F_X({\boldsymbol \omega})} ={ K} \frac{1}{F_Z({\boldsymbol \omega})}  \nabla {F_Z({\boldsymbol \omega})}
\end{equation}
Using $\nabla \log F_X({\boldsymbol \omega})=  \frac{1}{F_X({\boldsymbol \omega})}  \nabla F_X({\boldsymbol \omega})$,
 \begin{equation}
 \label{veceq2}
\nabla \log {F_X({\boldsymbol \omega})}=  (  I-  K)^{-1} { K}  \nabla \log {F_Z({\boldsymbol \omega})}
\end{equation}
 Note that  (see eg. \cite{kailath_book})
\begin{equation}
 K=  R_X ( R_X+  R_Z)^{-1}
\end{equation}
hence we have
\begin{align}
( I- K)=& ( R_X+  R_Z) ( R_X+  R_Z)^{-1} -  R_X ( R_X+  R_Z)^{-1}\nonumber \\
=&  R_Z ( R_X+  R_Z)^{-1}
\end{align}
and
\begin{align}
\label{apeq}
&( I- K)^{-1} K= [ R_Z ( R_X+  R_Z)^{-1}] ^{-1}  R_X ( R_X+  R_Z)^{-1}  \nonumber \\
&= [ R_Z ( R_X+  R_Z)^{-1}] ^{-1} [ R_X+  R_Z-  R_Z] ( R_X+  R_Z)^{-1}  \nonumber \\
&=  [ R_Z ( R_X+  R_Z)^{-1}] ^{-1} - I  \nonumber \\
&=  [ ( R_X+  R_Z)  R_Z^{-1}] - I   \nonumber  \\
&=  R_X  R_Z^{-1}+  I- I  \nonumber  \\
&= R_X  R_Z^{-1}
\end{align} 
 plugging (\ref{apeq}) into (\ref{veceq2}) we obtain,
\begin{equation}
\nabla \log {F_X({\boldsymbol \omega})}= { R_X}{ R_Z}^{-1}  \nabla \log {F_Z({\boldsymbol \omega})}
\end{equation}
Using the eigen decomposition of  $ R_X{ R_Z}^{-1} = U^{-1} \Lambda  U$ where $\Lambda $ is diagonal with eigen values $\lambda_1,...,\lambda_n$, we obtain
\begin{equation}
\label{lastvector}
{ U} \nabla \log {F_X({\boldsymbol \omega})}= \Lambda{  U} \nabla \log {F_Z({\boldsymbol \omega})}
\end{equation}
Similar to the scalar case, we can show the converse by retracing the steps in the derivation of the necessity. Note that none of these steps, (\ref{veceq})-(\ref{lastvector}), introduce any loss of generality, hence retracing back from (\ref{lastvector}) to (\ref{veceq}), we show that if (\ref{lastvector}) is satisfied, the optimal estimator is linear. 

\section{Proof of Lemma \ref{ax}}
\label{appd}
By the chain rule we have,
\begin{align}
\frac {\partial f( A \boldsymbol x)}{\partial x_i}&=\sum_{k=1}^n  \frac{\partial f( A \boldsymbol x)}{\partial [{ A\boldsymbol x}]_k} \frac{\partial [{ A \boldsymbol x}]_k}{\partial [{\boldsymbol  x}]_i}\\
&=\sum_{k=1}^n  \frac{\partial f( A \boldsymbol x)}{\partial [{ A\boldsymbol x}]_k} \frac{\partial ( {[A]_k}^T \boldsymbol x)}{\partial [{\boldsymbol x}]_i} \\
&=\sum_{k=1}^n  \frac{\partial f( A \boldsymbol x)}{\partial [{ A \boldsymbol x}]_k} [A]_{ki} \\
&=\sum_{k=1}^n {\partial_k f({  A\boldsymbol x})} [A]_{ki} \\
&= {[A]_i}^T \nabla f ({ A \boldsymbol x})
\label{son}
\end{align}
It follows from (\ref{son}) that $\nabla_x f ({  A\boldsymbol x})={ A}^T \nabla f({ A\boldsymbol x})$.


\section*{Acknowledgment}
This work is supported by the NSF under the grants CCF-0728986, CCF-1016861 and CCF 1118075. The authors would like to thank the editor and the referees for their comments that helped to clarify the results.

\ifCLASSOPTIONcaptionsoff
  \newpage
\fi

\bibliographystyle{IEEEbib}
\bibliography{ref}
\newpage
\begin{IEEEbiographynophoto}{Emrah Akyol}(S'03) received the B.Sc. degree in 2003 from Bilkent University (Turkey), the M.Sc. degree in 2005 from Koc University (Turkey), and the Ph.D. degree in 2011 in electrical and computer engineering from the University of California at Santa Barbara. From 2006 to 2007, he held positions at Hewlett-Packard Laboratories and NTT Docomo Laboratories, both in Palo Alto, where he worked on topics in video compression.

Currently, Dr. Akyol is a postdoctoral researcher in the Department of Electrical and Computer Engineering, University of California at Santa Barbara. His research focuses on source and source-channel coding, energy efficient communications, multimedia compression and networking, and the connections between estimation theory and information theory.
\end{IEEEbiographynophoto}

\vspace{-5in}

\begin{IEEEbiographynophoto}{Kumar B. Viswanatha}(S'08) received his B.Tech in electrical engineering in 2008 from the Indian Institute of Technology - Madras (IIT - Madras), Chennai, India and his MS in electrical and computer engineering in 2009 from University of California at Santa Barbara (UCSB), USA. He is currently pursuing his PhD in electrical and computer engineering at UCSB. He was an intern associate in the equity volatility desk at Goldman Sachs Co., New York, USA. His research interests include multi-user information theory, joint compression and routing for networks and distributed compression for large scale sensor networks.
\end{IEEEbiographynophoto}

\vspace{-5in}
\begin{IEEEbiographynophoto}{Kenneth  Rose} (S'85-M'91-SM'01-F'03) received the Ph.D. degree in 1991 from the California Institute of Technology, Pasadena.

He then joined the Department of Electrical and Computer Engineering, University of California at Santa Barbara, where he is currently a Professor. His main research activities are in the areas of information theory and signal processing, and include rate-distortion theory, source and source-channel coding, audio and video coding and networking, pattern recognition, and non-convex optimization. He is interested in the relations between information theory, estimation theory, and statistical physics, and their potential impact on fundamental and practical problems in diverse disciplines.

Dr. Rose was co-recipient of the 1990 William R. Bennett Prize Paper Award of the IEEE Communications Society, as well as the 2004 and 2007 IEEE Signal Processing Society Best Paper Awards.

\end{IEEEbiographynophoto}
\end{document}